\documentclass[a4paper,11pt]{article}
\usepackage{jinstpub} 
\usepackage{siunitx}
\usepackage{graphicx}
\usepackage{caption}
\usepackage{subcaption}
\usepackage{booktabs}
\usepackage{url}
\usepackage{tikz}
 \usetikzlibrary {arrows.meta,automata,positioning,shadows} 

\DeclareSIUnit{\inch}{in}
\DeclareSIUnit{\ph}{photon}

\title{Scintillation Light Detection in Polycrystalline Diamond Using Single Photon Detectors}

\author{Niccol\`o Gallice,}
\author{Aleksey Bolotnikov,}
\author{Erik M. Muller}
\author{and Thomas Tsang}

\affiliation{Brookhaven National Laboratory, Instrumentation Department\\
PO 5000, Upton, NY 11973, USA}

\emailAdd{ngallice@bnl.gov}

\abstract{This study investigates the scintillation properties of polycrystalline diamond for particle detection applications, particularly in neutron and alpha radiation environments. Polycrystalline diamonds provide a cost-effective alternative to monocrystalline diamonds while retaining essential detection properties. Photoluminescence measurements were performed to analyze emission spectra, revealing distinct characteristics based on impurity content and crystallinity. Scintillation responses were assessed using Silicon Photomultipliers (SiPMs), demonstrating the capability of polycrystalline diamond powders to respond to alpha irradiation, albeit with reduced resolution compared to traditional scintillators. A prototype neutron detector was developed by combining diamond powder with neutron-sensitive ${}^6$LiF, and its performance was evaluated through experimental testing and Geant4 simulations. The findings indicate that polycrystalline diamond-based detectors can achieve significant detection efficiency while remaining insensitive to gamma radiation, offering potential for portable neutron detection applications.}


\begin{document}
\maketitle
\flushbottom

\section{Introduction}
Diamonds have long been valued for particle detection due to their superior semiconductor properties, including a high band gap that ensures low dark current at ambient temperature and excellent radiation hardness. Solid-state detectors using diamonds have been developed to measure deposited energy via ionization charge collection \cite{bohonDevelopmentDiamondbasedXray2010,zouProtonRadiationEffects2020,bergonzoRadiationDetectionDevices2003,kaganDiamondRadiationDetectors2005}. However, achieving high efficiency and resolution requires monocrystalline diamonds with high crystallinity and purity, which are costly and size-limited.

Beyond charge-based detection, diamonds also exhibit scintillation properties \cite{1960PPS....76..670D, Nam1989DetectionON}. Historically, their use was constrained to natural diamonds with limited property control. Photoluminescence, associated with vacancies and dopants in the diamond lattice \cite{Photoluminescence_2017, photolumin_2017}, provides insights into scintillation mechanisms. Advances in Chemical Vapor Deposition (CVD) technology now allow the fabrication of large-area diamonds with controlled doping.

Recent studies \cite{UMEMOTO2023168789} have primarily focused on single-crystal diamonds, which face size limitations. This work explores scintillation in polycrystalline diamonds across various formats. Measurements are conducted using Silicon Photomultipliers, which are compact, rugged, and efficient, facilitating the development of portable diamond-based scintillation detectors.

This paper presents photoluminescence measurements of various diamonds in Section \ref{sec:photoluminescence} and their response to alpha irradiation in Section \ref{sec:scintillation_response}. The concept of a thermal neutron detector is introduced in Section \ref{sec:Neutron_detector} followed by details of the Geant4 simulation for optimizing detector geometry and preliminary results from thermal neutron irradiation and a comparison with a ${}^3$He detector.

\section{Photoluminescence}\label{sec:photoluminescence}

Emission spectra from diamonds depend on the type of impurities embedded in the lattice, as well as the vacancies and crystallinity of the diamonds. To understand the emission profiles, various samples were tested using a YAG laser with a wavelength of \SI{266}{\nano\meter} and pulse duration of \SI{20}{\pico\second}. The response was then measured using an Ocean Optics spectrometer, which has a sensitivity range from \SI{195}{\nano\meter} to \SI{995}{\nano\meter}.

The tested samples consist of high-purity diamonds and diamond powders. All samples exhibit a predominant peak in the \SIrange{400}{600}{\nano\meter} region. The diamond powders, however, show an additional infrared shoulder and a double-peaked structure in the visible range. The result suggests that high-purity diamonds have lower luminescence yields due to the absence of defects and impurities that act as luminescence centers, whereas diamond powders exhibit higher yields due to their lower purity.

In most samples, the emission peak lies within the blue-green region, where commercial silicon photomultipliers (SiPM) typically achieve efficiencies greater than \SI{30}{\percent}, with potential efficiencies reaching up to \SI{50}{\percent} \cite{HPK_S14161,LOIZZO2024169751}.

\begin{figure}[htbp]
     \centering
     \begin{subfigure}[htb]{0.45\textwidth}
         \centering
         \includegraphics[width=0.8\textwidth]{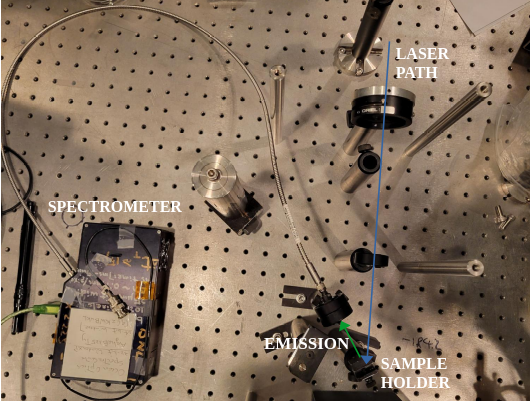}
         \caption{}
         \label{fig:photolum_setup}
     \end{subfigure}
     \hfill
     \begin{subfigure}[htb]{0.45\textwidth}
         \centering
         \includegraphics[width=\textwidth, trim={0.8cm, 0.3cm, 1cm, 0.8cm}, clip]{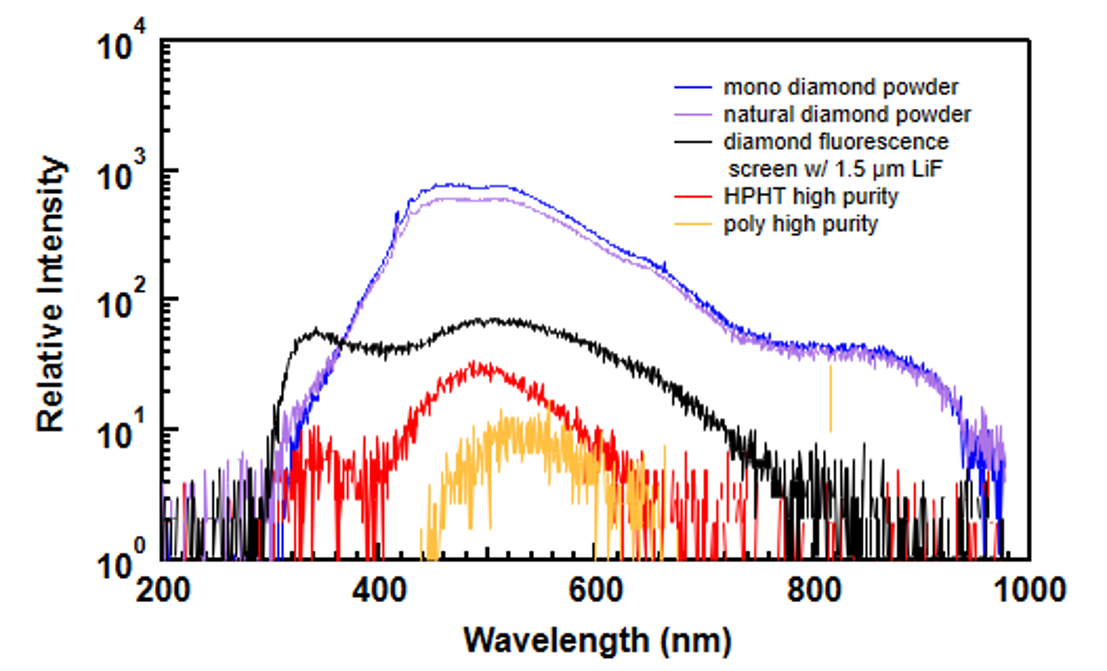}
         \caption{}
         \label{fig:photolum_spectra}
     \end{subfigure}
        \caption{On the left a picture of the setup used for photoluminescence characterization: a YAG laser beam impinges on the sample, and the luminescence response is guided with an optical fiber to a spectrometer. On the right the emission spectra from different samples.}
        \label{fig:photolum}
\end{figure}

\section{Scintillation response}\label{sec:scintillation_response}
To study diamond scintillation response, various samples were prepared: four different diamond powders on a double-sided tape that is placed on top of a glass substrate, and one CVD high-purity monocristal diamond with \SI{1}{\inch} diameter. The samples are prepared by laying the tape on the glass substrate and inserting them into a container with diamond powder, later shaking it to create a subtle layer of diamonds that sticks to the tape. The samples containing powders were labeled from L1 to L4 as they have different size or quality diamonds. As a reference sample, an LSO crystal was employed.

\begin{figure}[htbp]
     \centering
     \begin{subfigure}[htb]{0.39\textwidth}
         \centering
         \includegraphics[width=\textwidth]{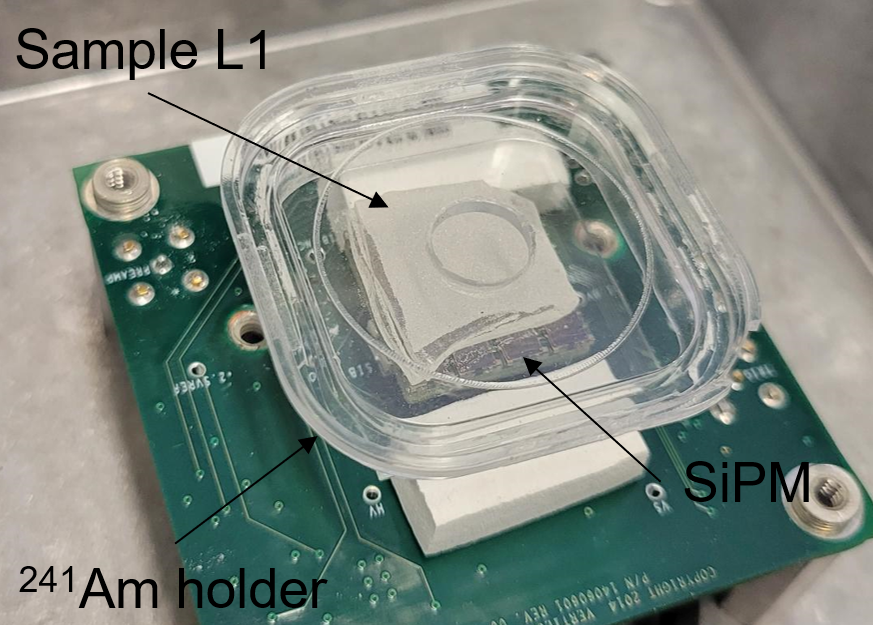}
         \caption{}
         \label{fig:scint_setup}
     \end{subfigure}
     \hfill
     \begin{subfigure}[htb]{0.6\textwidth}
         \centering
         \includegraphics[width=0.8\textwidth, trim={2cm 1.5cm 6cm 3cm}, clip]{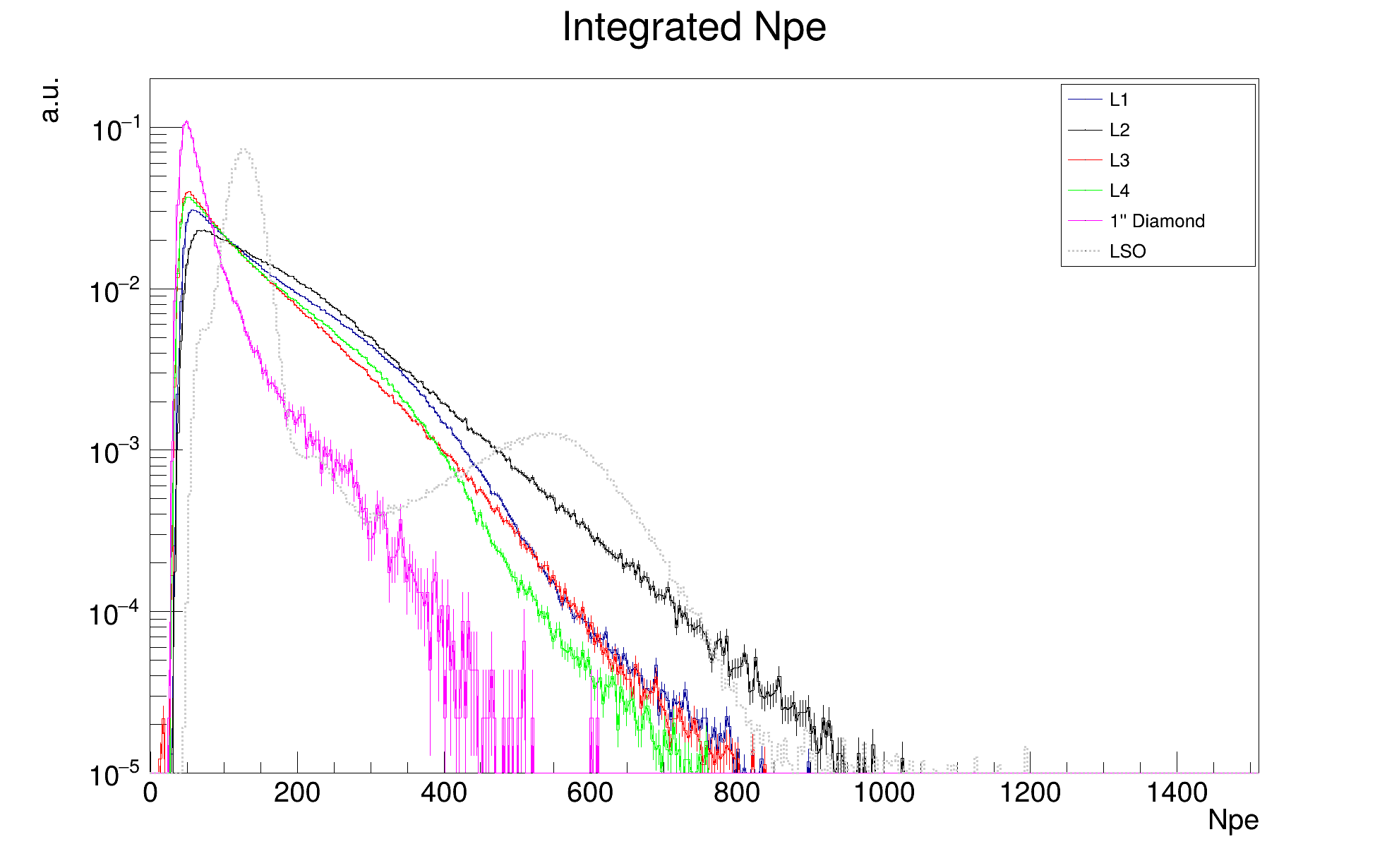}
         \caption{}
         \label{fig:scint_spectra}
     \end{subfigure}
        \caption{On the left (a) a picture of the setup used for scintillation characterization. On the right the spectra from various diamond samples and LSO control sample.}
        \label{fig:scint}
\end{figure}

The samples were positioned on a 4-by-4 matrix of Silicon PhotoMultipliers (SiPMs) (Figure~\ref{fig:scint_setup}). The sensors, part of the Onsemi C-series, feature a \SI{3}{\milli\meter} side length and a \SI{35}{\micro\meter} micro-cell size \cite{Onsemi_C-series}. The SiPM tile is connected to a Vertilon SIB616 front-end board, which provides a bias voltage of \SI{28.5}{\V}, and interfaces with a Vertilon IQSP580 data acquisition system.

The trigger is configured based on the sum of all 16 channels, with a threshold set just above the sensor's dark count level. The samples are irradiated using a ${}^{241}$Am alpha source, and the charge is recorded for each SiPM upon every triggered event. Each recorded entry from an individual channel is divided by the gain to determine the number of equivalent photoelectrons (PE), which are then histogrammed to construct spectra.

Each SiPM undergoes calibration using a pulsed blue laser at \SI{420}{\nano\meter}, delivering a light signal at each trigger. The charge collected by each SiPM is histogrammed and fitted to a multi-Gaussian function to determine the gain of a single photoelectron.

The spectra are illustrated in Figure \ref{fig:scint_spectra}. The gray line represents the control LSO sample's response, peaking around 550 PE. Conversely, the diamond samples do not exhibit a noticeable peak. This phenomenon is often observed in powder scintillators, where scattering, absorption, and inhomogeneities result in poor resolution. Nonetheless, all diamond samples demonstrate a scintillation response to alpha particles. The CVD 1-\si{\inch} diamond shows a lower yield, consistent with its high purity, whereas the powders display a larger response with a bump structure around 300 PE.

\section{Neutron detector concept, prototype and measurements}\label{sec:Neutron_detector}

Diamonds are primarily insensitive to gamma radiation due to their low atomic number, making diamond-based neutron detectors inherently immune to gamma-ray backgrounds. In this work, we explore the potential of developing a thermal neutron detector by combining diamond powder with a neutron conversion material, ${}^6$LiF. In this design, thermal neutrons are absorbed by ${}^6$Li nuclei, triggering a ${}^6$Li(n,$\alpha$)t reaction. This reaction produces an alpha particle with \SI{2.73}{\MeV} and a triton with \SI{2.05}{\MeV} of energy. These particles release energy in the surrounding diamond material, which scintillates. The resulting photons can be collected with a solid-state photodetector to record the event.

\begin{figure}[htbp]
\centering
    \begin{minipage}[c]{.38\linewidth}
    \centering
            \begin{subfigure}[t]{.9\linewidth}
                \includegraphics[width=\textwidth, trim={22cm 0 21cm 0}, clip]{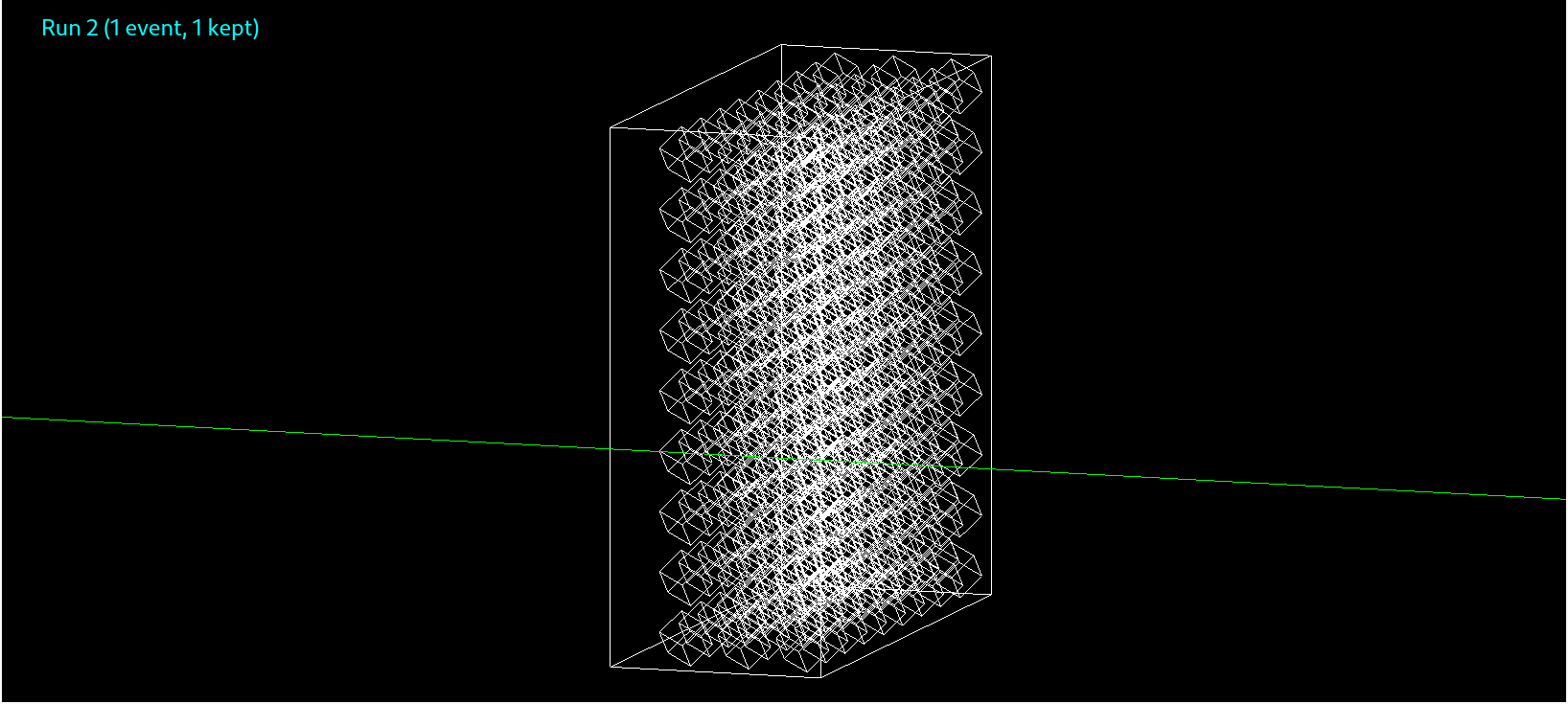}
                \caption{}
                \label{fig:G4_geometry}
            \end{subfigure}
    \end{minipage}
    \begin{minipage}[c]{.6\linewidth}
        \centering
        \begin{subfigure}[t]{\linewidth}
            \includegraphics[width=0.9\textwidth]{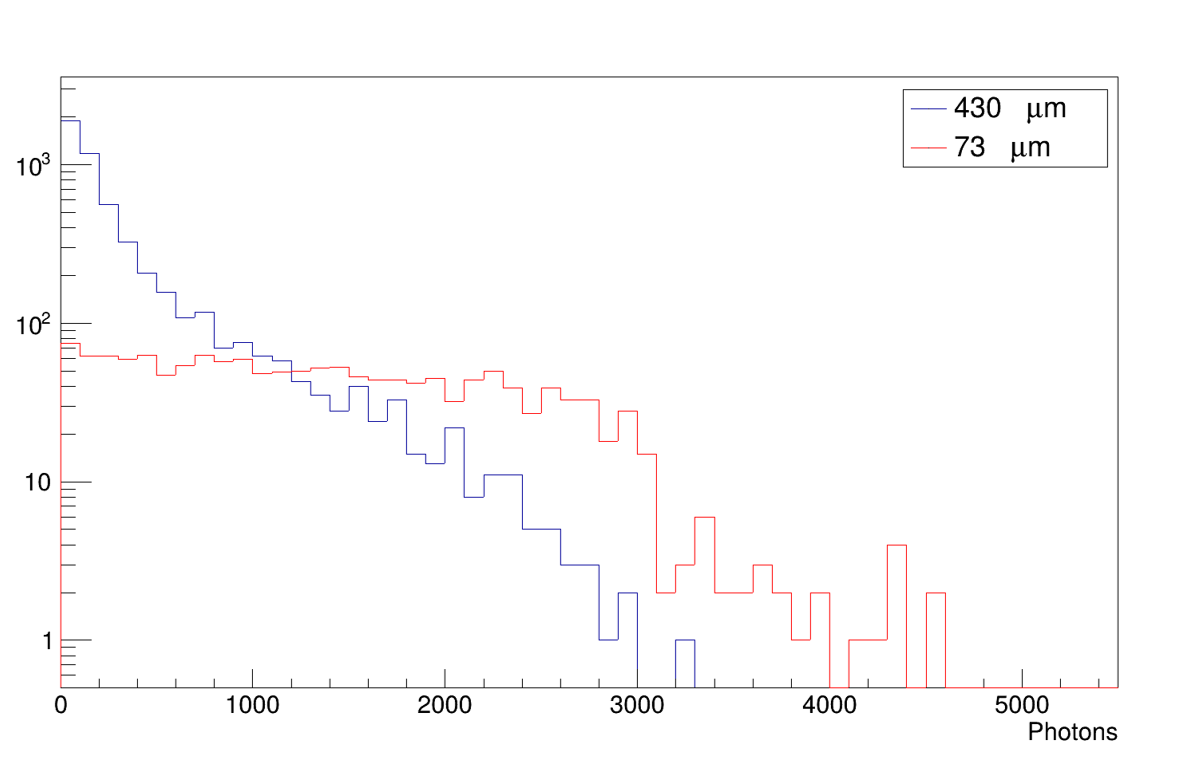}
            \caption{}
            \label{fig:G4_spectra}
        \end{subfigure} \\
        \begin{subfigure}[b]{\linewidth}
            \includegraphics[width=0.9\textwidth, trim={0 0.3cm 0 1.2cm}, clip]{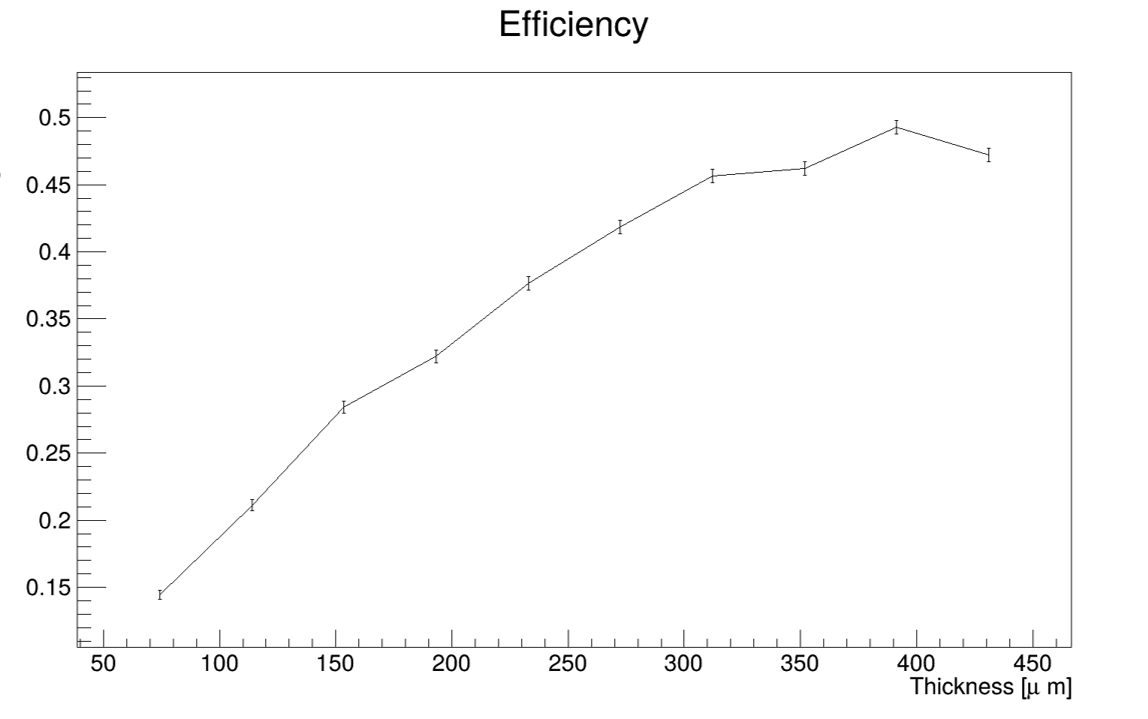}
            \caption{}
            \label{fig:G4_efficiency}
        \end{subfigure} 
    \end{minipage}
    \caption{In panel (a) the geometry used to simulate diamond powder in a $^6$LiF block. In panel (b) the spectra of photons detected by the sensor for two different thicknesses of the sample. In panel (c) the neutron detection efficiency as a function of detector thickness.}
    \label{fig:G4}
\end{figure}

A simulation of the detector concept was developed using Geant4~\cite{ALLISON2016186}. In this simulation, diamond grains are modeled as cubes with a \SI{20}{\micro\meter} side length, spaced between \SI{73}{\micro\meter} and \SI{430}{\micro\meter}, and embedded within a ${}^6$LiF cube that encloses them (Figure \ref{fig:G4_geometry}). To reduce computational expense, the cross-section is limited to \SI{0.3}{\milli\meter} $\times$ \SI{0.3}{\milli\meter}. The diamond material is modeled with a refractive index of \num{2.46} in the visible spectrum, an effective attenuation length of \SI{1}{\milli\meter}, and a light yield of \SI{3000}{\ph\per\MeV}. The ${}^6$LiF material is described with a refractive index of \num{1.4} and an attenuation length of \SI{500}{\micro\meter}. For optical photon simulation, a sensor surface with a refractive index of \num{1.5} is placed at the back of the ${}^6$LiF volume. A generator emitting thermal neutrons with \SI{125}{\meV} energy is set at \SI{1}{\cm} distance from the surface, with the other two coordinates uniformly distributed to avoid bias from the detector's anisotropy.

The thickness of the detector is varied between \SI{73}{\micro\meter} and \SI{450}{\micro\meter}, filling the volume with as many diamond grains as possible within the constraints. The photons detected at the sensor surface per event are histogrammed and presented in Figure~\ref{fig:G4_spectra}. The results show that energy resolution for the alpha and triton particles is lost, and that the average number of detected photons decreases with increasing detector thickness.

The detector efficiency is defined as the ratio of the number of detected neutrons to the number of generated neutrons. A neutron is considered detected if at least \num{20} photons are recorded by the sensor. The efficiency for each detector thickness is plotted in Figure~\ref{fig:G4_efficiency}. Based on conservative estimates, which will later be compared with direct measurements, it is possible to design a detector with approximately \SI{50}{\percent} efficiency for thermal neutron detection.

A prototype was created by mixing natural diamond powder with grains sized from \SIrange{30}{40}{\micro\meter} and ${}^6$LiF in water. This mixture was then applied onto a glass substrate using a brush: after the first layer is applied, it is allowed to dry, and a second layer is added on top. The sample consists of four layers. After drying, the sample is covered by an aluminum foil reflector and then positioned on the same sensors and acquisition system used for scintillation measurements. 

\begin{figure}[htbp]
    \centering
    \begin{subfigure}{0.45\linewidth}
        \begin{tikzpicture}
        \draw (0, 0) node[inner sep=0] {\includegraphics[width=\linewidth]{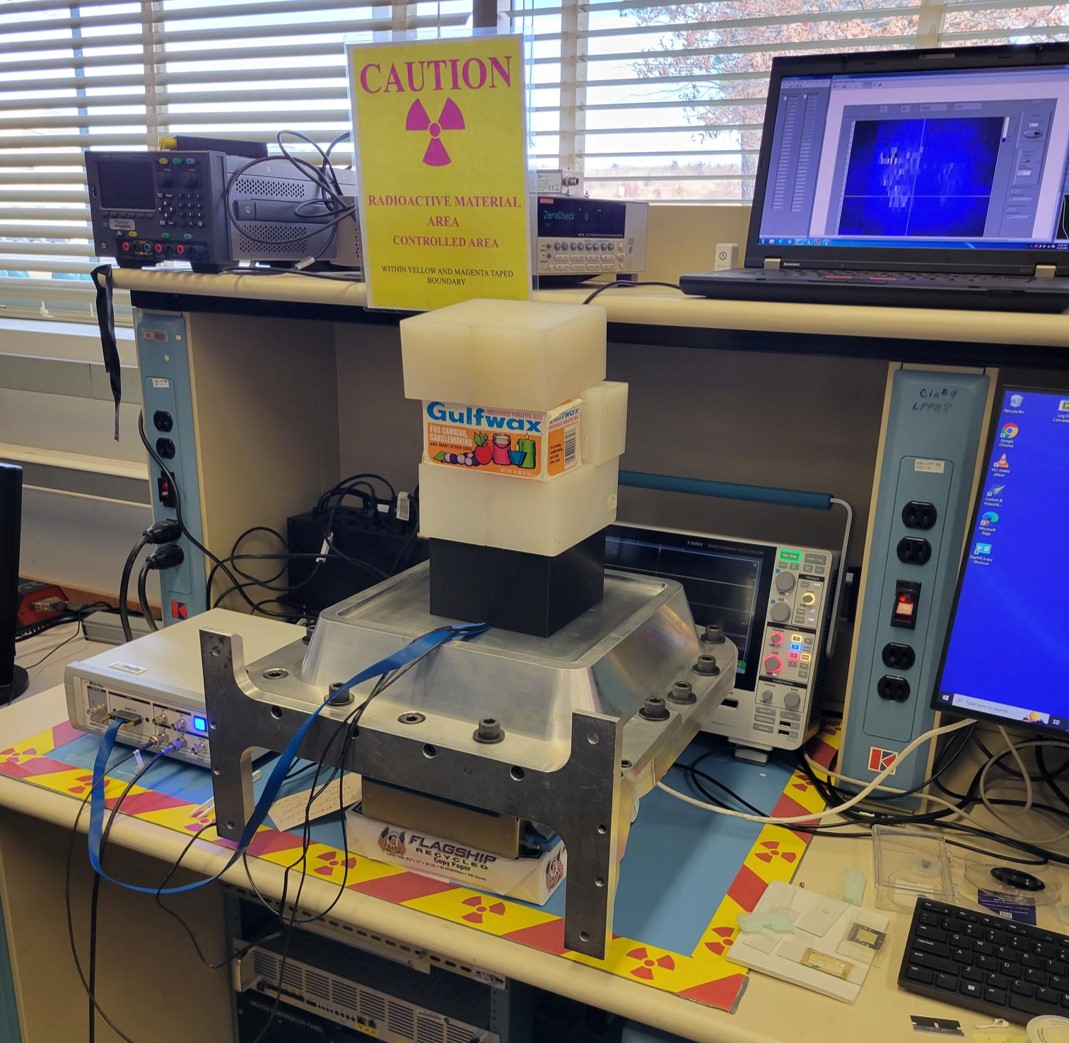}};
        \draw (2, 1) node {\textcolor{white}{$^{252}$Cf source}};
        \draw[-Stealth,white] (1,1)   -- (0,1);
        \draw (2, -0.2) node [text width=1.5cm] {\textcolor{white}{Diamond detector}};
        \draw[-Stealth,white] (1.2,-0.2)   -- (0.3,-0.4);
        \draw (2, -1.5) node {\textcolor{white}{${}^3$He detector}};
        \draw[-Stealth,white] (1,-1.5)   -- (0.3,-1.2);
        \end{tikzpicture}
        \caption{}
                \label{fig:n_setup}

    \end{subfigure}
    ~
    \begin{subfigure}{0.45\linewidth}
        \includegraphics[width=\linewidth]{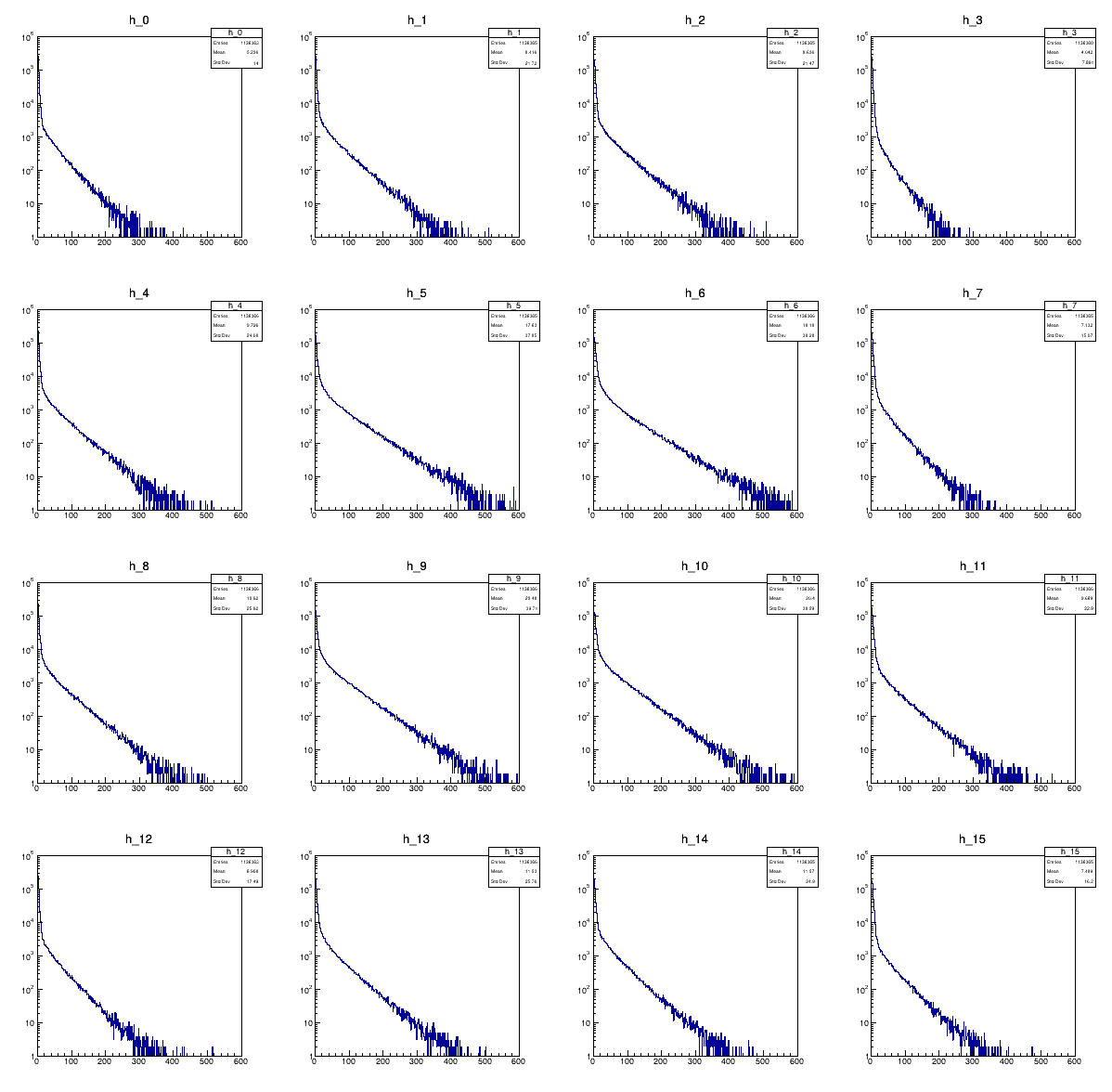}
        \caption{}
                \label{fig:n_channels}
    \end{subfigure}
    \caption{In panel (a) the neutron measurement setup. In panel (b) the photo-electron spectra from scintillation induced by neutron conversion to alpha and triton particles. The plots are arranged according to their physical location.}
    \label{fig:n_meas}
\end{figure}
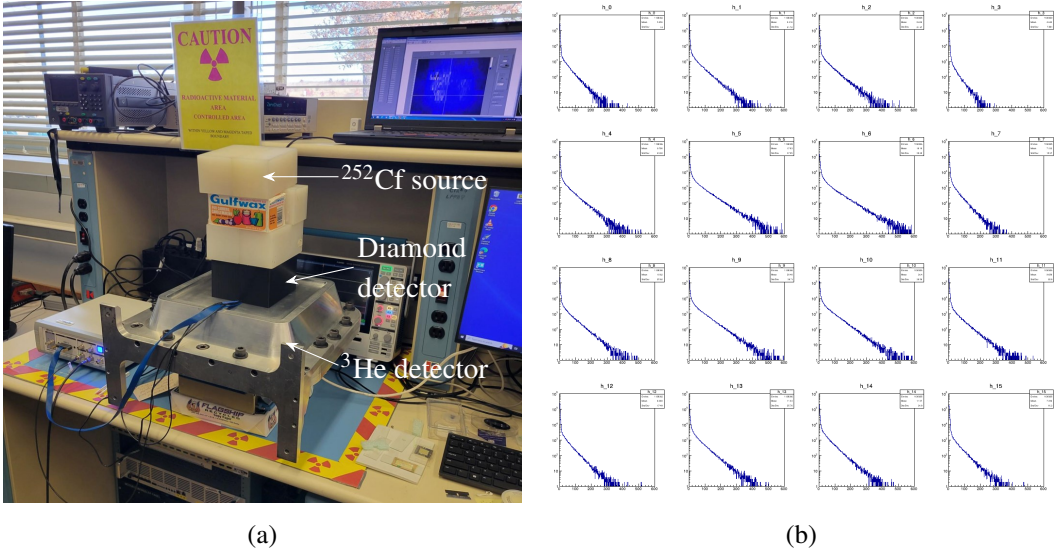

For these tests, the entire system is placed on a ${}^3$He detector developed at BNL \cite{BNL_He3}, as shown in Figure~\ref{fig:n_setup}. A ${}^{252}$Cf neutron source is placed on top of the two stacked detectors, surrounded by paraffin blocks to slow down the neutrons emitted by spontaneous fission to thermal energies.

Each of the two systems is triggered independently and operates on a separate data acquisition system. The data collected per channel from the diamond-based prototype is summarized in histograms shown in Figure~\ref{fig:n_channels}. These histograms confirm that scintillation light is generated by the sample when irradiated with neutrons. The light is produced by tritons and alpha particles resulting from neutron capture on ${}^6$Li nuclei.

The measured rate from the diamond-based prototype is \SI{0.44}{\hertz}, which can be compared to \SI{147}{\hertz} from the ${}^3$He detector. The effective surface areas of the two detectors are quite different: \SI{1.4}{\centi\meter\squared} for the diamond-based prototype and \SI{576}{\centi\meter\squared} for the ${}^3$He detector. Assuming a uniform neutron flux, a rough estimation of the efficiency ratio $\epsilon_\text{diamond}/\epsilon_{\text{He}-3}\sim 1.2$.

\section{Conclusions}\label{sec:Conclusions}
Using alpha particles emitted from the $^{241}$Am source, we measured strong signals from polycrystalline diamonds.   Photoluminescence spectra measured from different types of diamonds were found to be within a visible range, facilitating easy coupling with silicon photomultipliers. Using Geant4 simulations, we developed a concept for a thermal neutron detector with high detection efficiency. A prototype was built and compared to a large-area $^3$He neutron detector showing a comparable efficiency.  

\acknowledgments
We aknowledge DOE and Brookhaven National Laboratory for support of this work. We also thank BNL Instrumentation Department personnel for their precious insights, support and collaboration.

\bibliographystyle{JHEP}
\bibliography{biblio.bib}

\end{document}